\documentclass[11pt,titlepage,twocolumn]{article}

\usepackage{amsmath}

\usepackage{graphicx}

\usepackage[round,numbers,sort&compress]{natbib}
\usepackage{hyperref}
\DeclareTextSymbol{\degre}{OT1}{23} 
\newcommand{\latin}[1]{\textit{#1}} 
\usepackage{xcolor} 
\newcommand{\corr}[1]{#1} 
\newcommand{\corrbis}[1]{#1} 
\newcommand{\corrter}[1]{#1} 

\title{Visualization of \corrter{membrane loss} during the shrinkage of giant vesicles under electropulsation}
\author{Thomas Portet\\
	Institut de Pharmacologie et de Biologie Structurale, \\
         CNRS UMR 5089, Universit\'e Paul Sabatier, Toulouse, France\\
           and \\
   Laboratoire de Physique Th\'eorique, \\
	CNRS UMR 5152, Universit\'e Paul Sabatier, Toulouse, France. \\
	\and  Franc Camps i Febrer \\
	Institut de Pharmacologie et de Biologie Structurale, \\
         CNRS UMR 5089, Universit\'e Paul Sabatier, Toulouse, France.\\
         \and Jean-Michel Escoffre \\
         Institut de Pharmacologie et de Biologie Structurale, \\
         CNRS UMR 5089,  Universit\'e Paul Sabatier, Toulouse, France.\\
          \and Cyril Favard \\
        Institut Fresnel, \\
        CNRS UMR 6133, Marseille, France.
       \and Marie-Pierre Rols \\
        Institut de Pharmacologie et de Biologie Structurale, \\
         CNRS UMR 5089,   Universit\'e Paul Sabatier, Toulouse, France.
         \and David S. Dean \thanks{Corresponding author. Address:
           Laboratoire de Physique Th\'eorique,
	   CNRS UMR 5152, Universit\'e Paul Sabatier,
	   118 Route de Narbonne,
	   Toulouse, 31062, France.
	   Tel.:~(0033)561556463, Fax:~(0033)561556065} \\
	Laboratoire de Physique Th\'eorique, \\
	CNRS UMR 5152, Universit\'e Paul Sabatier, Toulouse, France. }

\begin{document}

\maketitle

\abstract{We study the effect of  permeabilizing electric fields
  applied to two different types of giant unilamellar vesicles, the
  first formed from EggPC lipids and the second formed from DOPC
  lipids. Experiments on vesicles of both lipid types show a decrease
  in vesicle radius which is interpreted as being due to lipid loss
  during the permeabilization process. We show that the decrease in size
  can be qualitatively explained as a loss of lipid area which is proportional
  to the area of the vesicle which is permeabilized. \corrter{Three possible modes of membrane loss were directly observed: pore formation, vesicle
  formation and tubule formation.}

}

\clearpage

%

\section*{ABBREVIATIONS - LIST OF SYMBOLS}
\begin{description}
\item[]EggPC: L-$\alpha$-Phosphatidylcholine (Egg, Chicken)
\item[]DOPC: 1,2-Dioleoyl-\textit{sn}-Glycero-3-Phosphocholine
\item[]DiIC$_{18}$: 1,1'-dioctadecyl-3,3,3',3'-tetramethylindocarbocyanine perchlorate
\item[]Rhodamine PE: L-$\alpha$-Phosphatidylethanolamine-N-(Egg Lissamine Rhodamine PE)
\item[]$E$: magnitude of the applied electric field
\item[]$\Delta\Psi$: transmembrane voltage
\item[]$\Delta\Psi_0$: initial transmembrane voltage induced by the first pulse at the poles of the liposomes.
\item[]$\Delta\Psi_c$: critical transmembrane voltage
\item[]$n$: number of pulses
\item[]$R(n)$: radius of the vesicle after $n$ pulses
\item[]$R_c$: critical radius
\item[]$W(n)$: rescaled radius of the vesicle after $n$ pulses $R(n)/R(0)$
\item[]$W_c$: rescaled critical radius $R_c/R(0)$
\item[]$\lambda$: fraction of the permeabilized area lost per pulse
\item[]$N_c$: number of pulses needed to enter the shrinking regime
\item[]$q$: probability that one pulse induces a transition from the pre-shrinking to the shrinking regime
\item[]SR: shrinking regime
\item[]PR: pre-shrinking regime
\item[]$a$: membrane thickness
\item[]$a_e$: membrane electrical thickness
\item[]$\epsilon_m$: membrane dielectric constant
\item[]$C$: constant depending on $R$, $a$ and the various conductivities of the problem
\item[]$\theta$: angle on the cell surface with respect to the direction of the applied field
\item[]$\theta_c$: critical angle
\item[]$A$: area of the vesicle
\item[]$A_p$: permeabilized area
\item[]$\Sigma_0$: initial surface tension
\item[]$\Sigma_{el}$: surface tension induced by the electric field
\item[]$\Sigma_{lys}$: lysis tension
\item[]$p$: dipole moment of the PC headgroup
\item[]$l$: length of the hydrocarbon chain
\item[]$\rho$: effective radius of the lipid hydrocarbon tail viewed as a cylinder
\item[]$\mu$: lipid tail hydrophobic free energy per unit of area
\end{description}

%

\section*{INTRODUCTION}

Electropermeabilization is a commonly used physical method where electric
pulses are applied to cells and vesicles and has been widely reviewed in the literature
\citep{Neu89,Wea95,Wea96,Tei05,Esc07,Fav07,Dim07}.
An effect, of major importance, of the electric
pulses is that under certain circumstances they can induce the transient
permeabilization of the cell plasma membrane. This permeabilization manifests
itself via the crossing of the cell membrane by molecules which would normally
not be able to permeate the cell membrane.  When subjected to
sufficiently large  electric fields,  vesicle membranes become
permeable to small molecules \citep{Neu72,Rol06} and flat membranes
show a marked increase in their electrical conductance \citep{Rob89}.
Small molecules appear to
cross permeabilized membranes via simple diffusion. However  complex
processes, such as electrophoresis and direct interactions with the
membrane,  come into play for larger molecules such as DNA.
Electropermeabilization is now regularly employed as a
delivery method for  a large variety of molecules such as drugs, antibodies,
oligonucleotides, RNA and DNA \citep{Rol98,Gol02,Ant05,Rol06, Fav07}.
Initial studies were carried out in vitro
on cells in culture, but as the technique has developed an increasing
amount of data has been obtained in vivo on
tissues \citep{Bel93,Geh99,Ser08} and the method is being adapted to the clinical context
\citep{Gil97,Got03}. Clearly the  method has a huge potential in the fields of cancer
treatment and gene therapy, offering, in some cases more efficient,
more controllable and safer treatment protocols (when compared to viral transfection methods for example).
From a purely  physical point of view
the  application of an electric field to a
lipid membrane has two notable  effects. The first is
a mechanical one where the stresses
caused by the field can deform the membrane, for instance causing a
spherical vesicle to deform into an \corr{ellipsoidal or cylindrical one  \citep{Win88,Kum91,Ris05,Ris06}}. This
deformation can be thoroughly understood in terms of a macroscopic continuum
description of the cell membrane in terms of its bulk electrical and mechanical
properties. The second  phenomenon
of electropermeabilization is much less well understood.
Despite its increasing popularity as a therapeutic method,
there are still many open questions about the underlying physical mechanisms
involved in electropermeabilization.  Indeed, at the simplest level, the
basic structural changes induced by the field on the membrane structure
are still to be fully understood. A number of physical theories have been
put forward to explain the phenomenon of electropermeabilization. Historically
the first explanations of electropermeabilization were based on classical
continuum theories which predict dielectric breakdown of the membrane at
a critical field strength \citep{Cro73,Abi79,Pas79,Nee89,Isa98,Sen02}.
The main problem with such theories is that
while predicting a dielectric/mechanical breakdown transition they do
not provide a description of the physical state of the permeabilized membrane.
Currently, the most popular explanation for electropermeabilization is that
pores are formed due to a local increase in the surface tension due to
the electric field \citep{Pow86,Gla88,Bar91,Neu03,Neu03b}.
This increase in surface tension energetically favors the
formation of pores which is otherwise energetically defavored by their line
tension, a similar theory was first introduced to explain the rupture of soap
films \citep{Der62}. In this theory the pores can become stabilized in a hydrophobic to
hydrophilic pore transition via  the rearrangement of the lipids at the pore
edges. Because the permeabilization is explained by the formation of pores, the
phenomenon described by this theory is referred to as electroporation. Recently numerical simulations
have confirmed that pores can be induced by strong electric fields
\citep{Tie04,Leo04,Tar05,Hu05a,Hu05b,Tie06,Woh06}, typically the systems simulated are small and no significant lipid
loss during pore formation has been reported.

 When discussing the phenomenon of electropermeabilization we must distinguish
between two key stages of the process (i) the physical change induced in the
membrane by the field (in the absence of molecules to be transported) and
(ii) the interaction of the molecules that are to be transported with the
modified membrane. At the simplest level, combination of steps (i) and (ii)
can be observed experimentally as a transport phenomenon using marked
molecules or via conductivity experiments. In this paper we demonstrate
that the step (i) can be indirectly detected via a change in the size of
giant liposomes under electropulsation and an associated direct visualization
of the expulsion of lipids from the liposomes. Concretely we study the
effect of a series of permeabilizing pulses,
well separated in time, on the size of giant unilamellar vesicles (GUVs).
In the experiments the radius of the GUV is measured after each pulse and
we find that each GUV studied shows, on average, a decrease in its
radius down to a critical radius beyond which its size no longer
changes. This decrease in size points to the fact that, during the
physical processes leading to electropermeabilization, lipids are
lost from the vesicle thus leading to a reduction in their size.

Our experiments are  not a direct study of permeabilization, however they
constitute an  indirect method of studying electropermeabilization
which is relatively straightforward to carry out and interpret in terms
of simple physical models which are relatively well
established. From an experimental point of view the crucial advantage of
using GUVs is that their composition can be varied and controlled and also their
membrane is not subjected to internal mechanical constraints as is the case for
living cells with cellular cytoskeletons.
Furthermore, their size is similar to that of mammalian cells, which allows a direct
visualization by an optical microscope.

Naively lipid loss during electropermeabilization
seems normal as if, for instance, pores are formed the lipids near the
edges of these pores will be subject to strong variations in the local
electric potential and the electric field. Charges and dipole moments
on lipids will interact strongly with the electric field and
variations of the electric field respectively. The forces involved may well
be  capable of tearing lipids from the membrane structure. However our
experiments suggest that the mechanism of lipid loss is a collective one
which involves the formation of small structures such as tubules and vesicles
as well as pores. A simple comparison  of electrostatic (dipole electric
field interaction) energy and hydrophobic free energy suggests that individual
lipids cannot be removed from the membrane.

The phenomenon of lipid loss due to an applied field has
previously been studied in
\citep{Tek01} but from quite a different point of view (in this study the
effect of single pulses was examined). DOPC vesicles
of sizes of the order of 20 $\mu m$ were subjected to pulsed electric
fields of the order of 1 $kV/cm$ and duration 700 $\mu s$. The
vesicles were observed using a standard fluorescent microscope and at
the cathode facing side single pores of the size of about 7 $\mu m$
were observed. Such pores were however seldom found on the anode
facing side. However it was inferred that this side was also
permeabilized but that the pores responsible were too small to be
observed. In the experiments it was also noted that up to 14\% of the
vesicle surface could be lost during the process of pore
formation/permeabilization.

%

\section*{EXPERIMENTAL SETUP}
\corr{We decided to work with two different \corrter{lipids}. However, we wanted phospholipids with identical head groups in order to obtain the same dipole behaviour.
Thus we used DOPC and EggPC, purchased from Avanti Polar Lipids (Alabaster, AL).}
The formation \corr{medium} is an
aqueous solution with 240 $mM$ \corr{sucrose}.
\corr{  The pulsation buffer is
an aqueous solution of 260 $mM$ glucose that also contains 1 $mM$ phosphate buffer $KH_2PO_4/K_2HPO_4$ (Merck, Darmstadt, Germany) in order to impose a physiological pH of 7.4, and 1 $mM$ sodium chloride (Prolabo, Briare, France) in order to achieve an
electrical conductivity in the range of a few hundreds of $\mu S/cm$. Conductivities of internal and external solutions are measured with a HI 8820 conductimeter (Hanna Instruments, Lingolsheim, France), and have the values $\sigma_i \approx 15 \ \mu S/cm$ and $\sigma_e \approx 460 \ \mu S/cm$ respectively. The osmolarities are 285 $mOsm/kg$ for the formation medium, and 305 $mOsm/kg$ for the pulsation buffer. These measurements were performed with an Osmomat 030 osmometer (Gonotec, Berlin, Germany).}
The different refractive indexes of the internal and external media yields
a contrast which enables the vesicles to be visualized using a
microscope, and the density difference allows the sedimentation of the
vesicles on the bottom of the chamber, thus reducing their distance
from the objective. EggPC liposomes are visualized by phase contrast,
and DOPC liposomes by fluorescence microscopy. We worked with two
different dyes \corr{(Rhodamine PE (Avanti Polar Lipids) and $\mbox{DiIC}_{18}$ (Molecular Probes, Eugene, OR)
) without any noticeable change in our experimental results}.
The vesicle formation method employed here is electroformation, as
described in \citep{Ang86}. We chose this technique because it is
simple, easily reproducible and has a good yield. Furthermore, a large
amount of the produced vesicles is unilamellar, as demonstrated in
\citep{Rod05}.

\subsection*{Electroformation}
\subsubsection*{Lipid solution}
The lipids are diluted in chloroform, at a mass concentration of 0.5
$mg/mL$. For DOPC vesicles, the fluorescent probe is added at 0.005
$mg/mL$. This preparation and the following steps can be performed at
room temperature, because the gel phase/liquid phase transition
temperature of the lipids used is much lower.
\subsubsection*{Formation chamber}
The chamber is made of two glass layers covered with Indium Tin Oxide
to ensure the electrical conductivity of the surface. The two layers
are separated by an adhesive silicone joint of 1 $mm$ width. The
connection with the generator (AC Exact, model 128; Hillsboro, OR) is
maintained by two wires, each one soldered on a small copper strip
stuck on the ITO slide. Then, 15 $\mu L$ of lipid solution is
deposited on the conducting sides of the glass slides. 
\corrter{The deposition is carried out slowly and at constant rate in a chamber held at
4\degre$C$ to slowly evaporate the chloroform and then the slides are dried under vaccum 
for a couple of hours in order to entirely remove the remaining solvent molecules.}

Finally, the slides are
sealed together, and the chamber is filled with the formation \corr{medium}.
\subsubsection*{Voltage application}
We apply a sinusoidal voltage of 25 $mV$ peak to peak at 8 $Hz$. The
voltage is increased by 100 $mV$ steps every 5 minutes, up to a value
of 1225 $mV$. It is maintained under these conditions overnight. Next
we apply a square wave of same amplitude at 4 $Hz$ for one
hour in order to detach the liposomes from the slides.

\subsection*{Electropulsation}
\subsubsection*{Pulsation chamber}
The chamber where the GUVs are subjected to the electric field is
composed of a glass slide and a coverslip. Two
parallel copper strips \corr{of thickness 70 $\mu m$}  are stuck on the slide at a distance of 1 $cm$ apart. The
coverslip is then stuck onto the dispositive with heated parafilm. \corr{The
chamber is  1 $cm$ long (between electrodes), 2.6 $cm$ wide  (width of the coverslip) and 250 $\mu m$ high (value estimated via measurements with a microscope). We first introduce 60 $\mu L$ of pulsation buffer} between the slide and the coverslip, while taking care of
filling the whole chamber so as to ensure the conductivity of our
medium. Next we add 5 $\mu L$ of our GUV preparation. Capillarity
phenomena prevents the solution from leaking out of the chamber.

\corr{The electrode thickness is about the size of our biggest liposomes, which represents only a quarter of the chamber height. We could not \latin{a priori} be certain of the homogeneity of the field. However, we solved numerically Laplace's equation with finite element software Comsol Multiphysics (Comsol, Burlington, MA) for the case of our geometry. We found that the field was almost homogeneous in the bottom part of the chamber between the electrodes, and that the size and shape of the permeabilized area were not significantly different from that computed for a geometry with much bigger electrodes (data not shown).}

\subsubsection*{Pulsation method}
Electropulsation is carried out  using a CNRS cell electropulsator (Jouan, St-Herblain, France) which delivered square-wave electric pulses. An oscilloscope (Enertec, St-Etienne, France) is used to monitor the pulse shape and amplitude. 
The process of electropulsation is performed directly under the microscope. For the
phase contrast visualization we used an inverted epifluorescence
microscope Leica DM IRB (Leica Microsystems, Wetzlar, Germany) \corr{equipped with a Princeton RTE/CCD-1317-K/0 camera (Princeton Instruments, Trenton, NJ) and a 40$\times$ Leica phase contrast objective}, and an
inverted confocal microscope Zeiss LSM 510 (Carl Zeiss, Jena, Germany) \corr{with a 63$\times$ Zeiss objective}
for fluorescence imaging.
\corr{Excitation at 543 $nm$ was provided by a $HeNe$ laser, and emission filter was a 560 $nm$ longpass.}
\corr{The pulse duration was not set to a few hundreds of $\mu s$ as in \citep{Tek01,Ris05,Ris06}, but to 5 $ms$ because this value is commonly used for gene transfer protocols in mammalian cells
\citep{Rol06}.}
In most cases, we apply pulses at 0.5 $Hz$. However, we
sometimes have to interrupt the pulse train for a few seconds in order
to re-center the image on the liposome of interest. Indeed, the
observed vesicle does not always stay immobile.
\corr{It often experiences a
translational motion toward the positive electrode, because of which we sometimes have to modify the
centering. This displacement was always directed toward the anode, irrespective of the  net electric charge of the fluorescent probe we used (negative for Rhodamine PE and positive for $\mbox{DiIC}_{18}$).} 
\corrbis{As we will see later, the direction of this motion is coherent with the sign of the $\zeta$ potentials of the vesicles, which does not depend on the type of dye chosen.}
Due to the need  to re-center the image from time to time, the frequency of the pulses is not constant over a whole experiment, but we checked this did not affect our
results. The time delay between two consecutive pulses is of the same
order of magnitude, ranging from 2 seconds to a few tens of
seconds. This duration seems to be much longer than the time needed by
the vesicle to relax after one pulse, therefore it does not matter if
pulses are separated by 2 or 20 seconds.
\corr{Direct observation showed that vesicles were distorted rapidly after the pulse application, but as far as the eye could see there was no visible size or shape change between two consecutive pulses.}
The pulse amplitude is
chosen according to the rule $ED=(4/3) \Delta \Psi_0=Const$ (see
details later for this choice), where $E$ denotes the amplitude of the
electric field, and $D$ the initial diameter of the GUV. The constant
is chosen to be 1.7~$V$. This choice means that at the beginning of every
experiment the potential difference drop, $\Delta \Psi_0$, across the
GUV membrane at the poles facing the electrodes is theoretically
(see later) equal to about 1.3 $V$ - this value is well beyond the
value of 200 $mV$ typically  cited as the permeabilization
threshold for Chinese hamster ovary cells \citep{Tei93,Ga97}
and of the order of that cited for artificial vesicles and other cell types \citep{Tso91,Wea96,Nee89}.
In the pulsation chamber the
distance between the electrodes is 1 $cm$ and so the potential applied
between the electrodes is $1.7/D \ V$, where $D$ is measured in $cm$
or conveniently $17/D \ kV$ if we measure $D$ in $\mu m$. The idea
behind this large choice of initial transmembrane potential $\Delta
\Psi_0$ is that the field will initially permeabilize the membrane and
continue to do so till the vesicle size becomes significantly smaller
than the initial one. We note that our protocol yields initial transmembrane
potentials which are slightly lower, but of the same order as
those in the experiments of \citep{Tek01} (which varied between 1.4-2.5 $V$).

The experimental strategy is simple. We focus on a liposome and we
measure its initial diameter. We then tune the voltage amplitude
according to the rule described above, and we apply a pulse train
until the GUV does not shrink anymore. We acquire one image between
two consecutive pulses \corr{(about 1 $s$ after each pulse)}, so we are sure that the vesicle has
experienced an electric pulse between two consecutive values of the
diameters we measure. Image processing tasks are performed with ImageJ
(National Institutes of Health, Bethesda, MD).

\subsection*{$\zeta$ potentials measurements}
\corrbis{We measured the average $\zeta$ potentials of our GUVs on a Malvern Zetasizer 3000 HS (Malvern, Worcestershire, United Kingdom), using the following method. We diluted 1 $mL$ of the GUV solution obtained after electroformation in 2 $mL$ of a special buffer containing 240 $mM$ sucrose, 1.5 $mM$ phosphate buffer  and 1.5 $mM$ sodium chloride. Vesicles are thus suspended in a medium containing 1 $mM$ sodium chloride, 1 $mM$ phosphate buffer and 240 $mM$ sucrose. This composition is the same as that of our pulsation medium, except for the 260 $mM$ glucose replaced by 240 $mM$ sucrose in order to avoid sedimentation of the vesicles which would make the measurement impossible. We then split the 3 $mL$ into two samples, on which we performed two series of ten measurements each.}

%

\section*{THEORY}
The basic theory which explains electroporation is based on the
modeling of the vesicle electrode system in terms of a weakly
conductive cell membrane of conductivity denoted by $\sigma_m$ with
external and internal media of much higher conductivities denoted by
$\sigma_e$ and $\sigma_i$ respectively. We denote by $R$ the radius
of the vesicle which is assumed to be spherical and indeed stays
spherical throughout the experiments. In our
experiments $R$ lies typically between 10 and 100 $\mu m$. The
thickness of the vesicle membrane is denoted by $a$ and typically has
the value of 4 $nm$. In the steady state, which is achieved on time
scales much shorter than the time over which the pulse is applied, the
electric potential $\Psi$ obeys Laplace's equation and if $\theta$
denotes the angle on the cell surface with respect to the direction of
the applied field which is of magnitude $E$ then the potential drop
across the membrane at that point is given by (see
\citep{Neu89} for instance for a detailed derivation)
\begin{equation} \label{eqn:deltapsi}
   \Delta\Psi = -CRE\cos(\theta) ,
\end{equation}
where $C$ is a constant depending on $R$, $a$, and the various
conductivities of the problem. In the limits where $\sigma_m \ll
\sigma_i$, $\sigma_m \ll \sigma_e$ and $a \ll R$ the constant $C$
becomes very simple and takes the value $C=3/2$. For the parameters of
the experiments carried out here we are close to the limit where $C$
takes this limiting value. The most important point for our analysis
here is that $C$ is independent of $R$.  We thus find
that for a thin membrane the electric field inside the membrane and
normal to its surface, denoted by $E_n$, is given by
\begin{equation} \label{eqn:En}
   E_n(\theta) = \frac{CRE\cos(\theta)}{a} \ .
\end{equation}
Eq.~\ref{eqn:En} demonstrates that there is a huge amplification of
the externally applied field across the membrane. This huge electric
field internal to the membrane causes structural changes. Whether this
structural change corresponds to the formation of pores, dielectric
breakdown or the formation of defects or vesicles is still open to
debate. However in experiments where permeabilization is measured
either via conductivity measurements of planar membranes or by direct
optical observation of the entry of marker molecules a consensus
exists that permeabilization occurs locally in the membrane when the
magnitude of the potential drop across the membrane $\Delta\Psi$
exceeds a certain threshold $\Delta\Psi_c$ which is estimated to be of
the order of 0.25-1.0 $V$ \citep{Tei93,Ga97,Tso91,Wea96,Nee89}.  This corresponds to a field within the membrane
of about 50-250 000 $kV/m$ (for a membrane of thickness 4 $nm$). This
critical threshold is seemingly quite universal, being largely
independent of cell and vesicle composition. There is an alternative
though largely equivalent physical explanation of field induced breakdown
of the membrane. The effect of a local potential drop $\Delta\Psi$ across
the membrane is to induce a local electrical surface tension $\Sigma_{el}$
which can be computed via the Maxwell stress tensor and is given by
$\Sigma_{el} = \epsilon_m \Delta \Psi^2 a/2a_e^2$ where $\epsilon_m$ is
the dielectric constant of the membrane, $a$ is its thickness and $a_e$ its
electrical thickness \citep{Nee89,Dim07}. If the initial surface tension of the membrane is
$\Sigma_0$ then upon applying the field the total tension is $\Sigma=
\Sigma_0 + \Sigma_{el}$. The tension of rupture of a lipid membrane  is called
the lysis tension $\Sigma_{lys}$ and thus when the local tension
$\Sigma$ exceeds $\Sigma_{lys}$, we expect the membrane to be destabilized. This
formulation is strictly equivalent to the existence of a critical value
of the local electric field in the membrane at which breakdown will occur.
However in this formulation we see that $\Delta\Psi_c$ will depend on the initial
surface tension of the vesicle $\Sigma_0$. Indeed such a dependence on $\Sigma_0$ has been reported experimentally \citep{Ris05}.
In terms of the initial and lysis tension the critical potential is given by

\begin{equation} \label{eqn:d_psi_sigma}
\Delta\Psi_c = \sqrt{\frac{2a_e^2}{\epsilon_m a}\left(\Sigma_{lys} - \Sigma_0\right)},
\end{equation}
and thus we see that the value of the applied field required to {\em affect} the membrane will depend
on the initial tension of the vesicle. In our study we are interested in the mechanism of lipid loss
and the $\Delta\Psi_c$ that induces lipid loss does not necessarily correspond to that necessary to
induce permeabilization, however it is reasonable to expect that the two critical potentials
have the same order of magnitude. Studies of electropermeabilization phenomena show that
the critical potential depends on the duration of the applied pulse, the critical potential
being smaller for longer pulses \citep{Rol98}. This means that the underlying
physical mechanisms rely on  activated processes such as nucleation events for
first order phase transitions. This means that an applied pulse may have no effect with
some probability, this probability should decrease with the amplitude and duration of the pulse.
In our experimental set up the liposomes are visibly  under an  initial tension and we also expect that
there is some distribution of initial tensions even for vesicles of the same composition and similar
sizes. The critical potential for each vesicle should therefore be expected to vary.

As we are looking at
vesicles we can neglect any possible modification of the transmembrane
potential due to cellular activity and thus assume that it is given
purely by Eq.~\ref{eqn:deltapsi}. Assuming that the mechanical and electric membrane thickness
$a$  and $a_e$ remains constant, there is a critical transmembrane potential drop
beyond which the membrane becomes permeabilized or susceptible to lipid loss.
\corrbis{Clearly, at fixed electric field parameters (amplitude and duration),} a cell can no longer be permeabilized when its radius is smaller than a certain
critical radius $R_c$ beyond which no part of the cell is
permeabilized. We thus expect that the permeabilization and thus
vesicle shrinkage will stop once the vesicle has this critical
radius. The region where the magnitude of $\Delta\Psi$ is maximal is
clearly that facing the electrodes, corresponding to $\theta = 0$ and
$\theta = \pi$, and so these are the last points where the membrane is
permeabilizable. The value of $R_c$ is thus given by
\begin{equation} \label{eqn:deltapsic}
   \Delta\Psi_c = CER_c \ .
\end{equation}
If we are in the situation where $R > R_c$ then about the pole at
$\theta=0$ the region where $\theta$ is between 0 and $\theta_c$ is
permeabilized and $\theta_c$ is given by
\begin{equation} \label{eqn:thetac}
   \theta_c = \arccos \left( \frac{\Delta \Psi_c}{CRE} \right) \ .
\end{equation}
This region gives one half of the total permeabilized area of the
vesicle which we denote by $A_p$. We thus find that
\begin{equation} \label{eqn:Ap}
   \frac{1}{2} A_p = 2 \pi \int_0^{\theta_c} R^2 \sin (\theta ) d\theta = 2 \pi R^2 \left(1-\frac{R_c}{R}\right) \ .
\end{equation}

Now we consider how the area loss upon a pulsation can be related to
the physical parameters of the system. The simplest idea is to assume
that the area lost is simply proportional to the permeabilized
membrane area. This does not presuppose the mechanism of lipid loss - we
simply assume that in the region where the field exceeds the critical value
the membrane structure is altered. This alteration can be interpreted as
a form of dielectric breakdown and where it occurs we assume that
lipids can be effectively lost from the membrane surface.

If $n$ denotes the number of pulses, treating
$n$ as a continuous variable we can write that on average:
\begin{equation} \label{eqn:dAdn}
   \frac{dA}{dn} = - \lambda A_p \ ,
\end{equation}
that is to say the average area lost per pulse is simply proportional to the area where the
critical membrane potential (or equivalently surface tension) is exceeded.
Note that we should really use a discrete difference equation rather than the continuous one
above, however we have, numerically,  checked that  the difference behavior is insignificant when
compared to the typical experimental errors. Now if we assume that $\Delta\Psi_c$ remains
constant throughout the experiment,
Eq.~\ref{eqn:dAdn} can be solved using $A = 4 \pi R^2$  to obtain
\begin{equation} \label{eqn:Rn}
   R(n) = R_c + (R(0)-R_c)\exp \left(-\frac{\lambda}{2}n\right) \ .
\end{equation}
Thus we expect an exponential decay to the critical value of $R_c$ as
given by Eq.~\ref{eqn:deltapsic}.  If we define the dimensionless
variable
\begin{equation} \label{eqn:Wndef}
   W(n)=\frac{R(n)}{R(0)}  \ ,
\end{equation}
then $W(n)$ obeys
\begin{equation} \label{eqn:Wn}
   W(n)= W_c + (1-W_c)\exp \left(-\frac{\lambda}{2}n\right) \ ,
\end{equation}
and $W_c$ is the asymptotic value of $W$ after a large number of pulses have been applied and beyond which the vesicle is no longer permeabilizable ; it is given by
\begin{equation} \label{eqn:Wc1}
   W_c =\frac{R_c}{R(0)} = \frac{\Delta \Psi_c}{CER(0)} \ .
\end{equation}
Now in the experiments if we choose  to apply fields $E$ such that $ER(0)$ is constant, then
if  $ \Delta \Psi_c$ and $C$ are constant we find that
\begin{equation} \label{eqn:Wc2}
   W_c=\frac{\Delta \Psi_c}{\Delta \Psi_0} \ ,
\end{equation}
where $\Delta \Psi_0$ is the initial experimentally imposed potential
drop at the poles of the cells and is by construction (\latin{i.e.}
via the choice of $E$) the same for all vesicles. With this choice of
$E$ all plots of $W$ as a function of the number of pulses $n$ should
collapse onto the same curve if $\Delta\Psi_c$ remains constant during the experiment
and if it is the same for all vesicles. All plots will have $W(0) =1$ and should
attain the asymptotic value $W_c$ after the same characteristic number
of pulses (as we have assumed that $\lambda$ is independent of $R$).

\corrbis{We stress here that if $ER(0)$ is taken to be constant then the normal component of the electric field within the membrane is the same for every vesicle studied at the beginning of each experiment and thus, independently of any theory used to analyze the results, we are always looking at systems where the local electric fields in the membranes are the same.}

Clearly three sources of additional complexity are neglected in the above analysis
(i) the surface tension will fluctuate during the permeabilization/lipid loss process
(ii) the local electric field seen by the vesicles will fluctuate due to the presence of
other vesicles \citep{Pav02} \corrbis{(iii)  we shall see in the section on experimental results  that several  mechanisms can be involved in the process of lipid loss (pore, vesicle and tubule formation) and  clearly the choice of a single fitting parameter for lipid loss per permeabilized area $\lambda$ is 
another simplification. Indeed $\lambda$ should be interpreted as an  average area loss parameter
due to the (at least) three visualized mechanisms of lipid expulsion.} 

\corrbis{The initial surface tension (which will have some distribution about an average value)
will also play a role in the initiation of the permeabilization and  lipid loss process. To what extent the vesicle retains a memory of this initial tension is an important point. If after each pulse it had the same tension, then the distribution of the values of $W_c$ would be a direct reflection of this initial surface tension distribution. However it is likely that the tension will vary after each pulse and indeed that the 
tension is a dynamical variable. Our experimental results imply that the reduction of the radius
is due to expulsion of lipid from the main vesicle but that some expelled lipid is still in contact with
the main vesicle (as in the case of tubules). These attached lipids will constitute a reservoir which
will modify the effective surface tension of the main vesicle and this tension itself will evolve 
if the system has not had time to equilibriate between pulses. We conclude that in fitting the data 
with the simple model presented here we should find a scatter in the resulting values of $\lambda$
and $W_c$ due to the points (i), (ii) and (iii) mentioned above.}

%

\section*{EXPERIMENTAL RESULTS}

\subsubsection*{Observations and data fitting}
\begin{figure}
   \begin{center}
      \includegraphics*[width=4.8in]{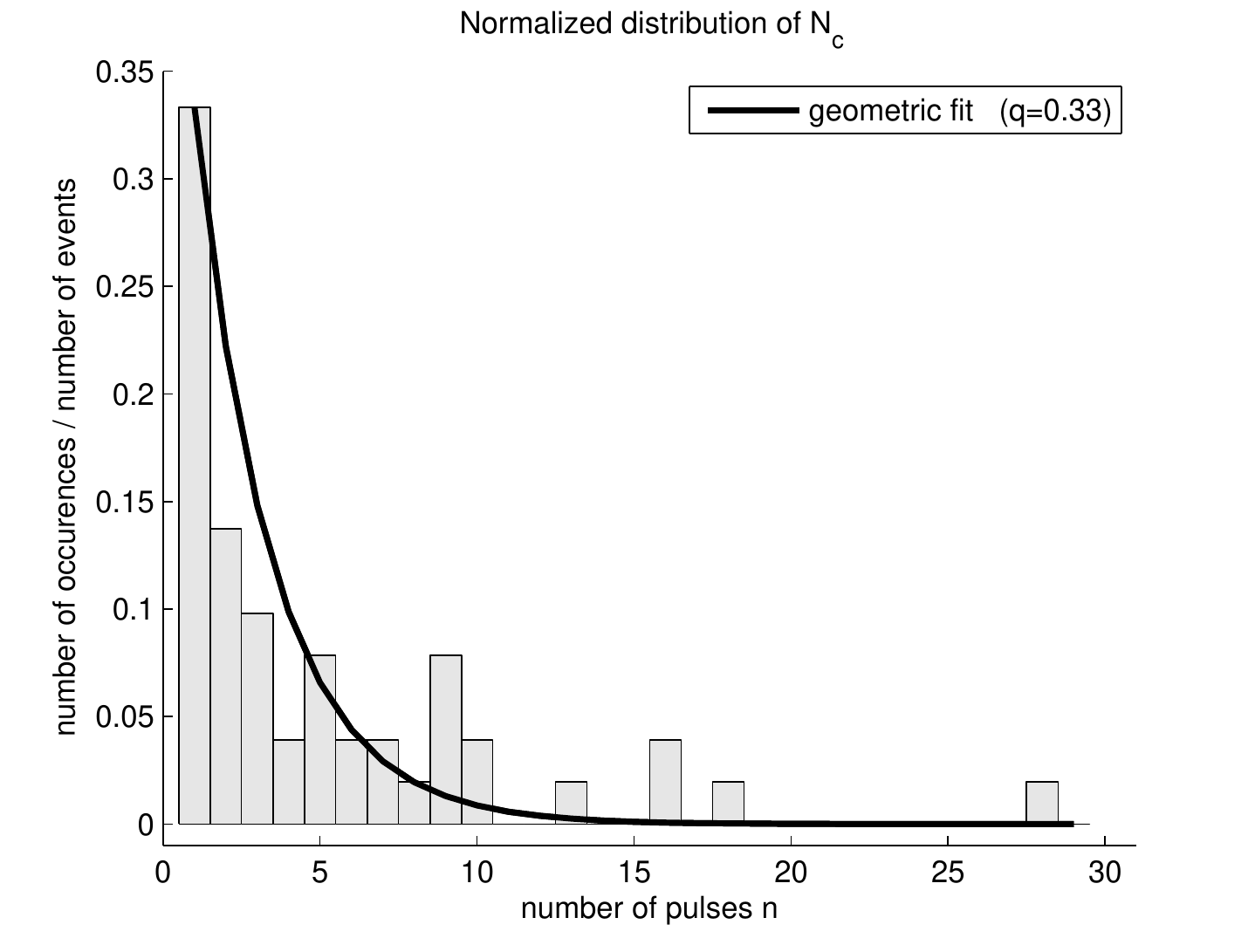}
      \caption{}
      \label{fig:Nc}
   \end{center}
\end{figure}

\corr{The existence of the critical radius $R_c$ was confirmed by the two following observations. (i) After a sufficiently large number of pulses had been applied, all the vesicles we could find in our sample had sizes lower than the one of the initial liposome of interest. (ii) We noticed that a liposome which had reached its critical radius could experience another shrinkage if the field magnitude was increased.}
We should mention that we sometimes saw vesicles disintegrating,
and thus we could not observe the size stabilization. We only kept
data corresponding to shrinking \emph{and} stabilizing GUVs, and we
finally gathered 51 data sets for DOPC and 47 or EggPC.  Another fact that must be mentioned is
the following. In some cases, the size diminution did not begin
immediately after the first pulse. We had to apply several electric
pulses before being able to detect radius decrease. A possible explanation for
this fact is that, like the permeabilization process, the mechanism for lipid
loss requires a change in the physical state of the membrane, the formation of
defects or pores for example. The effect of the field is therefore twofold -
it allows for the formation of defects and once defects are present the field
along with the presence of the defects allow for lipid loss. We may assume that
the creation of defects is an activated process and at each pulse the
membrane develops defects with some probability $q$. Note that we assume it is
only the defect creation process which has this probabilistic
nature (once the vesicle size has begun to decrease, lipids are
expelled after each pulse as long as the vesicle radius is greater
than $R_c$). In order to describe this phenomenon, we suppose that one
vesicle can be found either in a pre-shrinking (no defects) or in a shrinking
(with defects) regime (PR or SR respectively), the transition to the SR
after a pulse being a stochastic event occurring with constant probability $q$,
independent of the number of pulses applied before.  This hypothesis
of a random event is legitimate because our model should incorporate
the intrinsic stochastic nature of permeabilization processes \citep{Wea96}. The
fact that $q$ does not depend on $n$ is justified if we assume that a
vesicle having experienced a ``harmless'' pulse recovers the same
state it had in the PR.  Within this modified framework, the former
expression of the scaled variable $W(n)$ (Eq.~\ref{eqn:Wn}) now reads
\begin{equation} \label{eqn:WnNc}
W(n)= \mathcal{H}(N_c-n-1) + \mathcal{H}(n+1-N_c) \left[W_c^{fit} + (1-W_c^{fit})\exp \left(-\frac{\lambda^{fit}}{2}n\right)\right]
\end{equation}
where $\mathcal{H}$ denotes a Heaviside function taking the value 1 for
a positive argument and 0 otherwise, and $N_c$ the critical number of
pulses needed before entrance in the SR. This means that the fitted
curve will be constant up till $N_c$ and then decay exponentially after
$n=N_c$. We have denoted the critical value of $W_c$ given by the fit as
$W_c^{fit}$ and the effective value of $\lambda$ estimated from fitting is denoted by $\lambda^{fit}$. In terms of our theory we expect the average value of $W_c^{fit}$ to be
concentrated about $W_c$ with fluctuations about this value.
All fits were performed with
the formula given by Eq.~\ref{eqn:WnNc}, so we obtained values of
$N_c$, $W_c^{fit}$ and $\lambda^{fit}$ for each of the 51 DOPC data sets.  With
assumptions described above, the random variable $N_c$ should follow
a geometric (discrete and memory-less)  distribution.
We checked this by plotting the normalized histogram
of $N_c$, and as Fig.~\ref{fig:Nc} shows, the values of $N_c$ are well fitted by a geometric distribution of the form
\begin{equation}
\mbox{Probability}(N_c=n)=q(1-q)^{n-1} \ .
\label{eqn:geo}
\end{equation}
The shown fit yields the value $q=0.33$, which means that $N_c$ has the average value $\langle N_c \rangle=1/q=3$.
In Fig.~\ref{fig:data}, we
present four examples of data sets ($\times$ marks) and associated
fits (full lines). Diamond marks correspond to the images shown in
Figs.~\ref{fig:joli1} and~\ref{fig:joli2} depicting the different
mechanisms of lipid loss (see details below).  Except for liposome C
which immediately starts to shrink, we can clearly identify the PRs,
the SRs and the stabilization of sizes. 
\corrbis{Detailed information about pulse spacing for data from Fig.~\ref{fig:data}, which is not constant over a whole experiment because of the lateral motion of the vesicles, can be found in Table S1 in the Supplementary Material.}

\subsubsection*{Quantitative analysis - DOPC}

As a first step in our data analysis,  we can take the average of all the experimental curves and
then carry out a fit, this yields the values ${\lambda}=0.16$ and ${W_c}=0.65$.
The fit also yields the   number of pulses
 necessary to put the liposome in the active state, where lipid loss can be induced,
 to be ${N}_c = 1.73$.  The experimental data was also examined to see if there was any
 correlation between the fitted value of $W_c$ and $\lambda$ with the initial vesicle radius
 $R(0)$, no appreciable correlation was seen, thus validating our hypothesis that the vesicle
 shrinkage can be well described in terms of the rescaled (dimensionless) quantity $W(n)$.
 A second way to estimate the parameters of the model is to fit $\lambda$ and $W_c$ for each curve
 individually to obtain $\langle \lambda^{fit}\rangle$, $\langle W_c^{fit}\rangle$ and $\langle N_c\rangle$,
 the average value of the fitting parameters averaged over the individual experiments. The values obtained were $\langle \lambda^{fit}\rangle= 0.25$, $\langle W_c^{fit}\rangle = 0.58$ and $\langle N_c\rangle = 4.99$.
This value of $\langle N_c \rangle$ agrees well with that of 3 estimated by the geometric distribution fit to the histogram of the fitted values for $N_c$.

\begin{figure}
   \begin{center}
      \includegraphics*[width=4.8in]{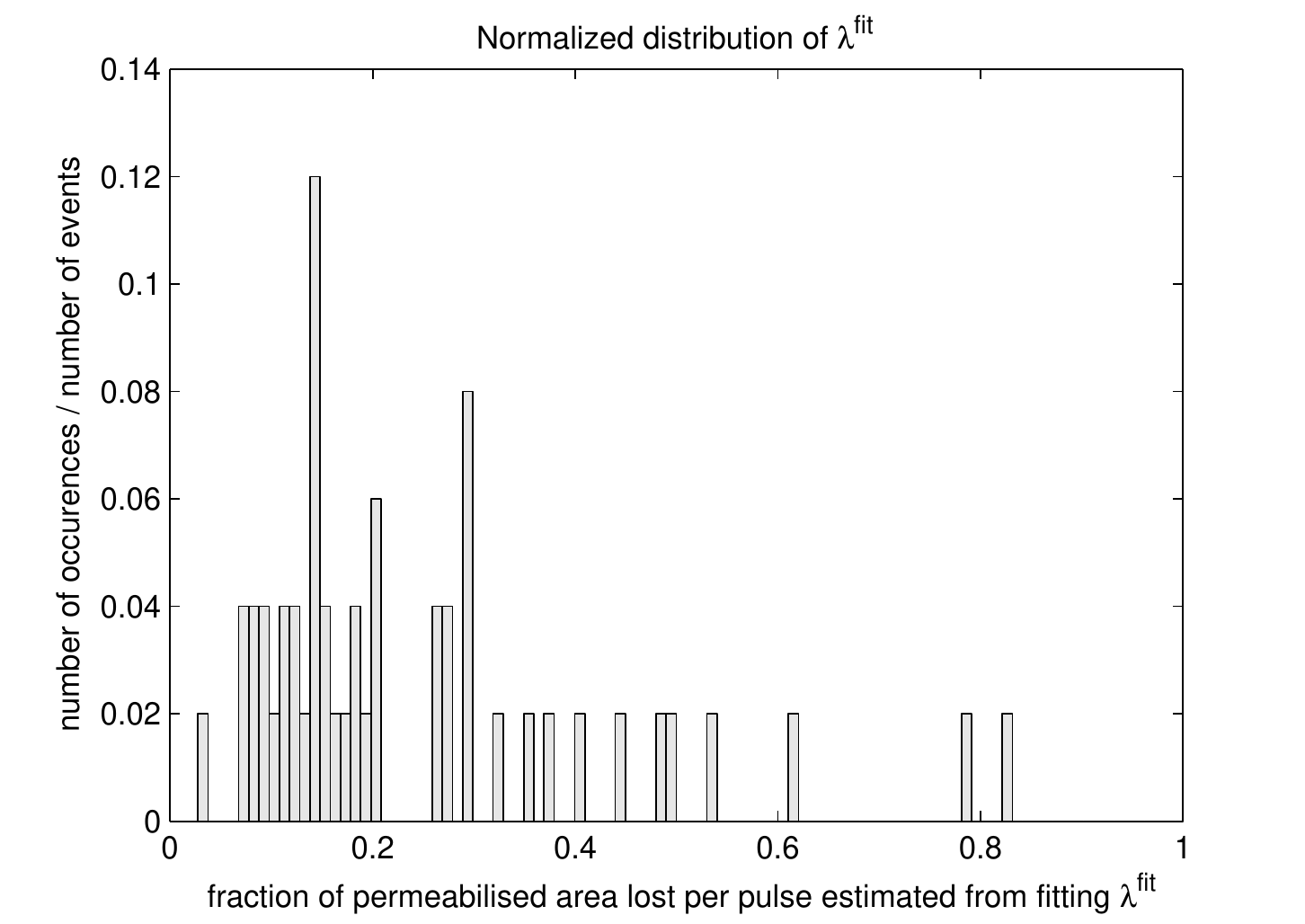}
      \caption{}
      \label{fig:lambda}
   \end{center}
\end{figure}

\begin{figure}
   \begin{center}
      \includegraphics*[width=4.8in]{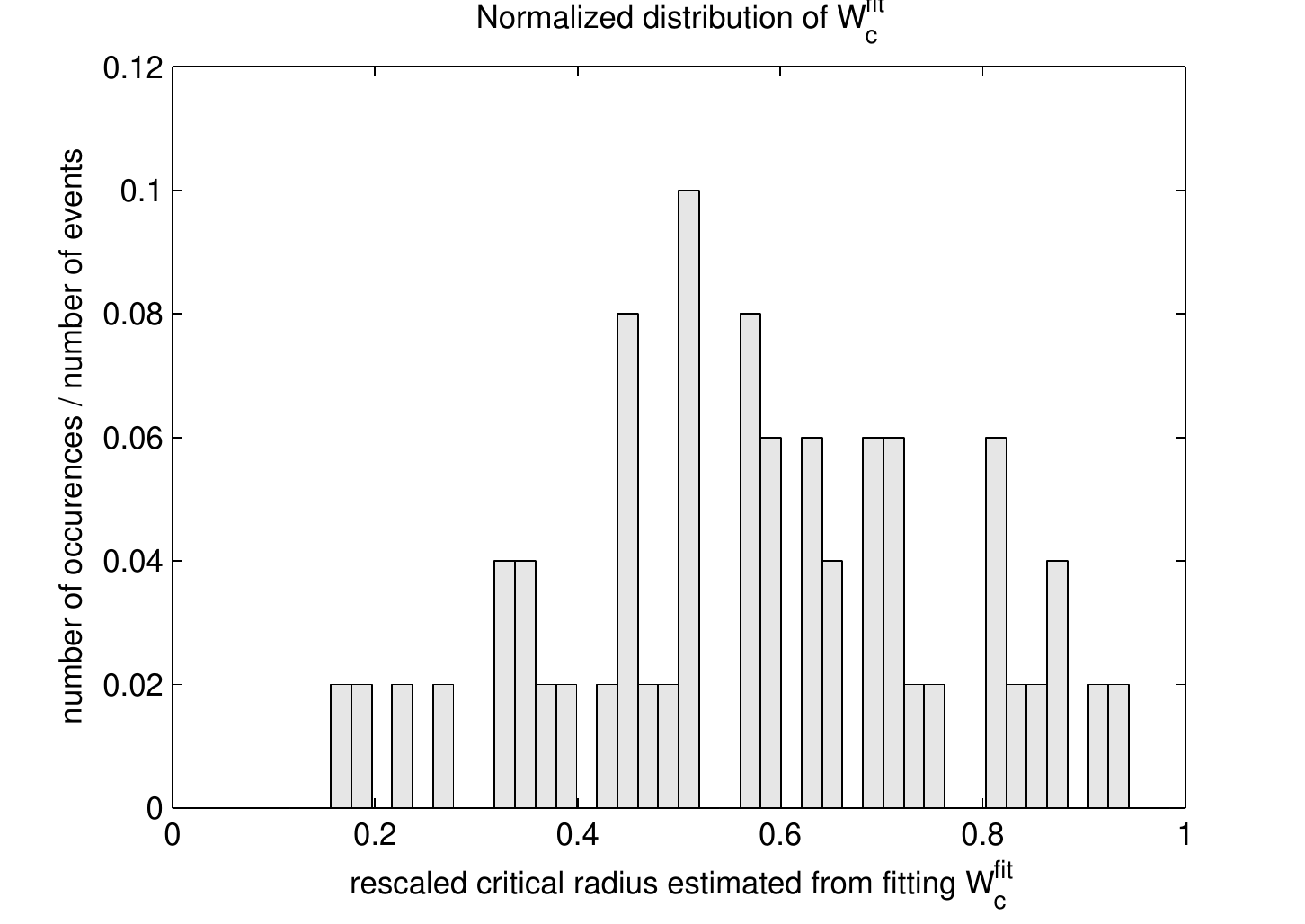}
      \caption{}
      \label{fig:Wc}
   \end{center}
\end{figure}

Figs.~\ref{fig:lambda} and \ref{fig:Wc} show
the histograms of $\lambda^{fit}$ and $W_c^{fit}$ respectively. 
\corrbis{As mentioned in the theory section, in fitting the data with our simple model we should expect
to see variation in the values of $\lambda$ and $W_c$ obtained due to fluctuations of the surface tension (both initial and during the permeabilization process), local electric field and possibly the effective number of defects created after the $N_c$ pulses needed to enter into the permeabilized state.  
We note that it  has indeed been demonstrated, in \citep{Ris05,Ris06}, that the critical potential necessary to induce permeabilization is indeed dependent on the surface tension.}

\subsubsection*{Quantitative analysis - EggPC}

The experiments with EggPC were performed first and at that time we had not yet made the considerations about the PR and the SR.
We only kept data sets corresponding to immediately shrinking vesicles, therefore in this section $N_c=1$ for each liposome.
Despite this simplification, we did the same data processing as that described for DOPC. The fit on the average of all experimental curves yields the values ${\lambda}=0.27$ and ${W_c}=0.77$. The values of the fitting parameters averaged over the individual experiments are $\langle \lambda^{fit}\rangle= 0.31$ and $\langle W_c^{fit}\rangle = 0.69$.

\subsubsection*{About the anode-directed motion of the vesicles}
\corrbis{The translational motion we observed was always directed toward the anode, suggesting that the GUVs could carry a net negative charge, even with a positively charged fluorescent dye. We checked this by measuring the $\zeta$ potential of the vesicles in a medium with ionic composition equivalent to that of our pulsation medium, the sugar composition being different to avoid vesicle sedimentation making the measure impossible. We examined four different types of vesicles: EggPC alone, DOPC alone, DOPC labeled with Rhodamine PE, and DOPC labeled with DiIC$_{18}$. We did not use EggPC vesicles with a fluorescent dye, because our experiments involving EggPC were performed via phase contrast microscopy, without any probe.}

\corrbis{For all four vesicle compositions, we find an average $\zeta$ potential of the order of $-20$ $mV$, a value in agreement with what was found in \citep{Car08} for DOPC GUVs, and whose sign is consistent with our observations. This corresponds to a negligible negative surface charge for the GUVs, less than one elementary charge per thousand of lipids. This residual electric charge possibly due to lipid impurities manifests itself only via the anode-directed motion of the vesicles, because of the large magnitude of the applied electric field.  
}

\subsubsection*{About the initial pH asymmetry}
\corrbis{Internal and external media of our GUVs were not at the same pH conditions (6.6 and 7.4 respectively). It was thus questionnable if this pH asymmetry had any significant influence in our experiments. The answer is no, based on the three following arguments. First, it is true that local pH gradients can induce the formation of tubular structures \cite{Kha08}. However, such gradients have a magnitude of $\sim$ 4 pH units, much higher than our 0.8 pH units.
Then, as can be seen on our phase contrast images (data not shown) GUVs become permeabilized during pulsation, and experience a mixing of their internal and external media. Thus the initial pH asymmetry should disappear after a few permeabilizing pulses.
Finally, the observation that the vesicles are stable and do not exhibit any shape changes until the electric field is applied corroborates the fact that the initial pH asymmetry of our GUVs has no significant effect.}

\subsubsection*{Mechanisms of lipid loss}

\corr{One of the most fascinating aspects of the experiments is the wide variety of mechanisms
of lipid loss that can be observed.
Three different mechanisms of lipid loss are observed when the lipids are fluorescently marked
as is the case on our experiments on DOPC liposomes (these observed mechanisms do not show any appreciable dependance on the probe employed). We emphasize here that the term
lipid loss implies loss of lipid from the bulk spherical part of the vesicle, the lipid ejected
appears in most cases to remain attached to or close to the parent vesicle.}

\begin{figure}
   \begin{center}
      \includegraphics*[width=4.8in]{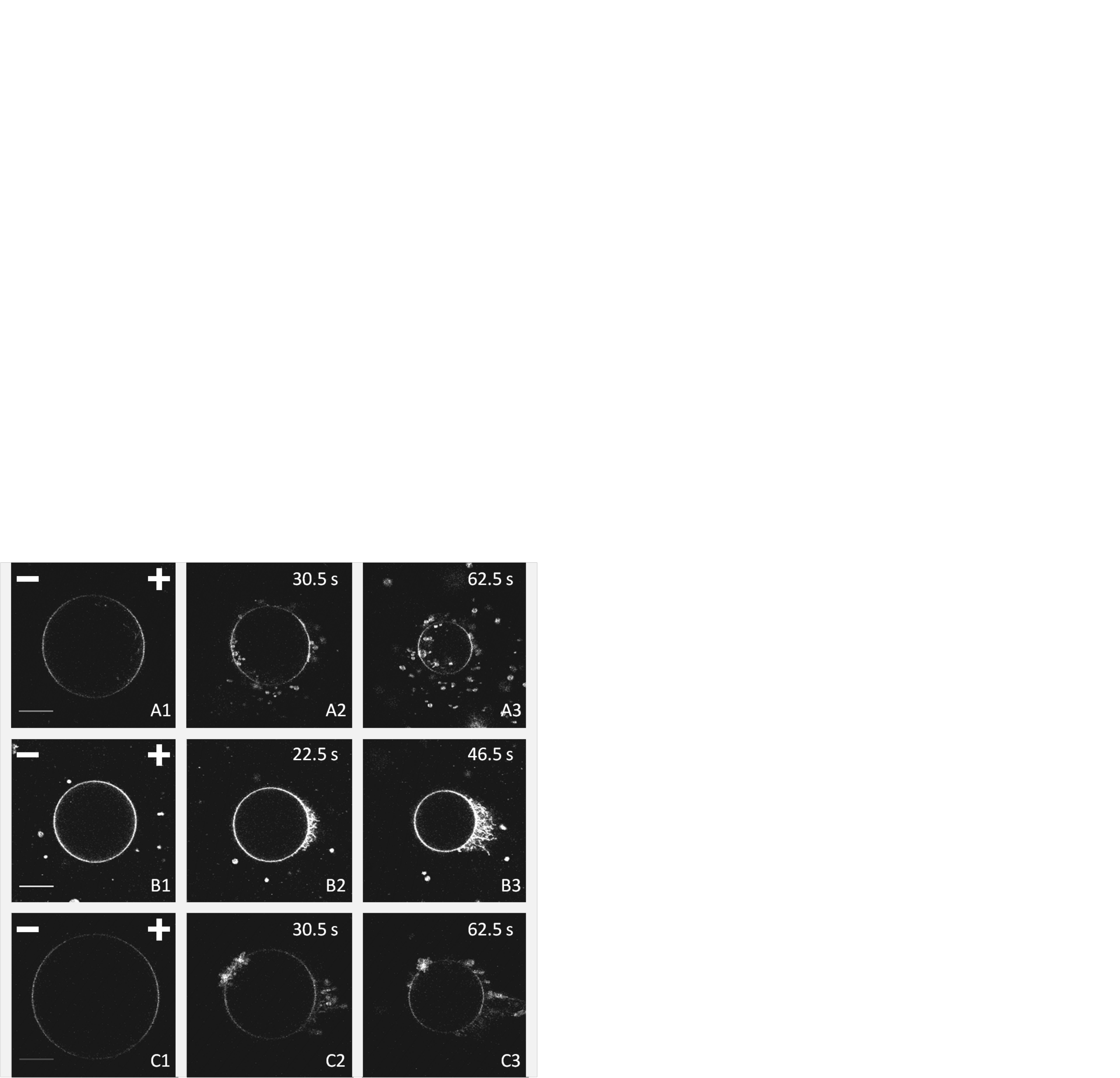}
      \caption{}
      \label{fig:joli1}
   \end{center}
\end{figure}

The three basic mechanisms are
shown in Figs.~\ref{fig:joli1} and~\ref{fig:joli2}. Images were
taken with the confocal microscope.

The first and most frequent
mechanism is the formation of small vesicles at both the anode and
cathode-facing poles. Those vesicles are mainly thrown out of the
GUVs, but some of them were also driven inside the GUVs. Liposomes A
and C of Fig.~\ref{fig:joli1} lost their lipids in such a manner \corrbis{(Movie S1 available online shows that mechanism for another GUV).} Interestingly a similar
phenomena has been reported when high frequency alternating electric fields are applied
to sea urchin eggs \citep{Mar95}, firstly the cell is deformed and   elongated by the
field and then this cell splits into two smaller cells and a number of much smaller vesicles.

The
second phenomenon we could observe (see photographs for liposome B in
Fig.~\ref{fig:joli1}) was the creation of lipid tubules on the
exterior of the anode-facing hemisphere \corrbis{(Movie S2 available online shows that mechanism for liposome B).} DOPC molecules expelled from the membrane
rearranged in the form of tubular structures, whose lengths grew with
the number of applied pulses. These structures initiated from the pole
facing the positive electrode and remained attached to the vesicle. However they then appeared to diffuse away from the pole towards the equator (while remaining attached to the membrane)
and appeared to cover most of the anode-facing hemisphere as shown in Fig.~\ref{fig:joli1}.
\corr{We  also saw on the cathode facing side of Fig.~\ref{fig:joli2}  that tubules can grow on the interior surface of the liposome. These structures also
diffuse toward the equatorial regions, the number and size of tubules however being smaller. This
mechanism of tubule formation appears to be stronger on the anode-facing hemisphere.}

Finally, we also noticed the presence of
pores on the cathode-facing hemisphere (as did \citep{Tek01}). This was a quite rare
observation, but it is normal because our acquisition times were of a
few hundreds of milliseconds, the same order of magnitude than the
lifetimes of such pores \citep{Tek01}. Liposome D, which has
entered the SR after 16 pulses, is found to have pores after 16 and 18
pulses, as show images D2 and D4 of Fig.~\ref{fig:joli2}. On the next
images, we can see the beginning of the formation of the tubular
structures described previously. We thus conclude that those two
mechanisms could occur \emph{together} for a same vesicle. The fact
that we detected only a few GUVs exhibiting pore formation is certainly
due to the too low acquisition speed of our experimental setup. Recently it
has been shown that pore formation can be induced in vesicles by solubilizing
the membrane \citep{Rod06} and that this process of pore formation is also
associated with membrane loss and thus vesicle shrinkage.
\corrbis{An animation of the shrinkage of liposome D associated with pores and tubules formation is available online in Movie S3.}

\corrbis{The eventual long term evolution  of the structures described above (after pulsation has been stopped) varied from one experiment to another. The small vesicles, in most cases, diffused away from the liposome and the vesicle radius stayed constant. However the behavior of the tubular structures exhibited wide variation. Some of the tubules broke away from the GUV and diffused away, sometimes forming vesicles and sometimes not. Other  tubules remained attached to the vesicles, exhibiting polymer-like fluctuations. In some cases they were  reabsorbed into the GUV membrane after a time of the order of minutes.  In fact,  the eventual fate of tubules   was strongly dependant on their environment, notably on whether other vesicles came in contact with them or not. In the cases where tubules were reabsorbed, the volume of the vesicle they were attached to increased and the final state 
of the  vesicle was often non-spherical and appeared to be under little tension (in agreement with
the idea that the attached lipids act as reservoir of lipid for the main vesicle).}

%

\section*{DISCUSSION}

Giant liposomes subjected to pulsed DC electric fields diminish in size and
lose lipids via several observable mechanisms, vesicle ejection, tubule formation
and pore formation. This is quite different to what is observed in living cells
which tend to swell under electroporation \citep{Kin77, Abi93,Abi94,Gol98}.
The experiments, along with the associated model, provide us with the following
picture of lipid loss due to applied pulses. The lipid loss proceeds by a two
stage process. First if the applied field is high enough a membrane passes
from an inactive state where it has no induced defects to one where defects
are present. We have seen that this process is of an exponential or character
reminiscent of radioactive decay. Secondly, for DOPC composed vesicles, once defects are present the
membrane loss per pulse is of the order of $\lambda\approx 0.20$ of the area in which the
transmembrane potential exceeds the critical value, denoted here by
$\Delta\Psi_c$. From our estimate $ W_c= 0.65$ obtained by fitting the average of all curves, we find that on average
$  \Delta\Psi_c =   W_c \times  \Delta\Psi_0 =0.65\times 1.3 \ V\approx 0.85 \ V$. If we use the average value of $\langle W_c^{fit}\rangle$ obtained by fitting the individual curves, we obtain $\Delta\Psi_c \approx 0.75 \ V$.

These values of  $\Delta\Psi_c$ are to be compared with those reported for certain cell
membranes $\Delta \Psi_c\approx 1 \ V$ \citep{Tso91,Wea96} and tension
free vesicles (1-stearoyl-2-oleoyl phospatidycholine and dioleoyl
phosphatidyglycerol) \citep{Nee89} where $\Delta \Psi_c \approx 1.1 \ V$. Similar results apply for
EggPC but in this case, $\lambda \approx 0.29$ and there is thus, with comparison to DOPC, more lipid loss per unit
area of where the critical transmembrane potential is exceeded. The estimated value of $W_c$ obtained by fitting the average of all curves  is 0.77, which gives a critical transmembrane voltage of  1 $V$. The estimate from the average values obtained over individual fits yields a value of 0.69 for $W_c$, thus leading to a critical transmembrane voltage $\Delta \Psi_c \approx 0.89 \ V$.

Recently numerical simulations have provided much
insight into the membrane organization occurring during the membrane
permeabilization process \citep{Tie04,Leo04,Tar05,Hu05a,Hu05b,Tie06,Woh06}.
The picture emerging is one where the strong
electric field present in the membrane causes water molecules (via
their dipole interaction with the applied field) to penetrate into the
membrane. There is an initial formation of so called hydrophobic pores
because the water molecules are in proximity to the hydrophobic core
of the membrane. Subsequently the lipid head dipoles re-orientate to
form hydrophilic pores where the lipid heads line the inside of the
pore. The mechanism behind this reorientation involves hydrophobic
effects and electrostatic effects. For example dipole moments which
are orientated normal to the membrane surface (which is roughly the
case for DOPC) are favorably aligned on one side of the membrane but
not on the other. This means that on the side where they are well
oriented the field keeps them straight toward the normal. However on
the side where they are mal-orientated they can lower their energy by
turning in toward the core of the membrane. This tendency to turn
inside the membrane lowers their electrostatic energy and aids the
formation of hydrophilic pores. The same effect is clearly present
before water penetration into the bilayer core and helps to form
defects which favor penetration by water molecules. This explains why
formation appears to be initiated from a particular membrane side in
electrically neutral membranes. However in numerical simulations lipid
loss from the membrane is not generally observed during pore formation
and pore resealing. This could be because the time scales over which
the simulations are carried out are too short. Indeed it is difficult
to see, if we accept the above image of the pore formation mechanism,
how lipid loss to the extent observed in our experiments can be
explained by such processes.  The main differences between the
experiments here and numerical simulations is that the system here is
much larger and that the pores formed are an order of magnitude larger
than those seen in simulations (which can be interpreted as
pre-pores). We have seen that vesicle formation seems to make a major
contribution to the observed lipid loss and there is presumably a
minimal size that a vesicle can have (for thermodynamic and mechanical
reasons), thus if the simulated system contains less lipids than
required to build a vesicle of minimal size then lipid loss by
vesiculization cannot be observed. Another possible mechanism for lipid loss is that lipid
head group dipoles which are mal-orientated, instead of turning into
the membrane to be better orientated are simply expelled from the
membrane. This expulsion will increase the free energy of the lipid
due to hydrophobic interactions but lower the electrostatic
energy. The hydrophobic component of the free energy increase could be
lowered by the formation of small vesicles into which these expelled
lipids could be incorporated. We recall that in smaller vesicles the
electrostatic energy of mal-orientated lipid head group dipoles is
much smaller due to the scaling with $R$, the vesicle radius, of the
potential drop across the membrane. The idea that single lipids can be
extracted due to the field turns out to be unrealistic. The dipole
moment $p$ of the PC head group is about 20 Debyes (see
\citep{Pas99} and references therein), this means that the
maximal electrostatic energy of a mal-oriented dipole is of the order
$E_D \approx p (\Delta \Psi / a)$, where $\Delta \Psi$ is the
potential drop across the membrane. However the hydrophobic energy of
a lipid tail placed in water is given by $E_{hydro} \approx 2 \pi \rho
l \mu$ where $l$ is the total length of the hydrocarbon chain and
$\rho$ is its effective radius (viewed as a cylinder). Clearly the tail
length is approximately related to the membrane thickness by $l
\approx a / 2$.  The term $\mu$ is a hydrophobic free energy per unit
of area and takes a value of about 40 $mJ/m^2$
\citep{Isr00}. The effective cylindrical radius of the lipid
hydrocarbon tail is estimated at 0.8 $nm$ (there is of course really
two tails each of radius approximately 0.4 $nm$
\citep{Isr00}).  Equating these two energies yields a critical
transmembrane potential beyond which lipids can be torn out directly
by the field as
\begin{equation} \label{eqn:deltapsistar}
  \Delta \Psi^{\star} \approx \frac{\pi a^2 \rho \mu}{p}\approx 24 \ V \ .
\end{equation}
This value of $\Delta \Psi^{\star}$ is to be compared with the value
given typically for the critical potential drop across the membrane
necessary to achieve permeabilization which, as previously mentioned,
is about 200 $mV$ for a wide range of membrane types. In addition the
electric field seen by the lipid heads is only the amplified one if we
assume that the head region is of low conductivity having a value
close to that cited for the total membrane conductivity.  We thus
conclude that for permeabilization seen in the range of voltages
of our experiments, a
simple mechanism of tearing out lipids is unlikely to occur (although
this mechanism could conceivably play a role when high intensity short
pulses are applied).  The conclusion of the above estimation is that
lipids must be ejected together in structures that minimize their hydrophobic
energy such as micelles, tubules and vesicles as is indeed seen in our experimental
results.

There is a clear asymmetry in our observations of lipid loss, in agreement with the observations
of \citep{Tek01} when we observed pore formation it was on the cathode facing side of the
liposome. However  the anode facing side was the one where the formation of tubules was favored.
The mechanism of symmetry breaking could well be related to the anisotropic dielectric structure
of the membrane due to the behavior of its lipid components.

Another interesting feature of our results is that the vesicle does not always lose
lipid material from the first pulse onwards. This implies that the vesicle needs to be in a
particular state (induced by the field with some probability) in order to enter into the
shrinking regime (SR). The difference between the SR and pre-shrinking regime is unclear,
but one could speculate that in the SR the membrane has defects which facilitate the loss of
lipids. {\corrbis The number and nature of defects created at the inception of the SR is presumably stochastic in nature and could be responsible for the variations in the parameter $\lambda$ seen in our experiments}.   The continued application of pulses then leads to a number of visible \corrter{modes of membrane loss}, vesicle formation, tubule formation and pore formation. In the context of applied DC pulses only
pore formation had been previously reported \citep{Tek01}. Vesicle formation due to alternating fields
has been reported \citep{Mar95} but the underlying physics appears quite different as in the
presence of AC fields the formation of small vesicles occurs via the fission of the initial cell into two similar sized daughter cells. Perhaps the most striking phenomenon is that of tubule formation which leads to a hair like structure of tubules around the liposome. Thus repeated application of short DC pulses leads to the shrinkage of artificial vesicles and a rich phenomenology of lipid structure formation.
As a final comment the phenomenon of lipid loss observed here seems to support aspects of the
phase transition model of electropermeabilization \citep{Sug79}.
In this model the electric field can induce a transition from a state where the bilayer is thermodynamically stable to one where smaller units, for example micelles, are thermodynamically preferred. The fact that the lipid loss process is not always immediately initiated, when $N_c\neq 1$, supports the first order
nature of the transition.

{\bf Acknowledgments:} We would like to thank Justin Teissi\'e for useful discussions on this work.
 We also acknowledge financial support from the {\em Association Fran\c caise contre les
 Myopathies} and the {\em Institut Universitaire de France}. We would like to thank \'Emile Perez and Plamen Kirilov from the 
 IMRCP in Toulouse for allowing us to use their facilities to measure the $\zeta$ potentials of our vesicles. Our group belongs to the CNRS consortium CellTiss.

%
%



%
%
%
\bibliography{bibliodd}

\clearpage
\section*{Figure Legends}

\subsubsection*{Figure~\ref{fig:Nc}.}
Normalized distribution of the values of $N_c$ obtained after fitting of experimental data for DOPC vesicles. Solid line is a fit to a geometric distribution of the form given in Eq.~\ref{eqn:geo}, yielding the value $q=0.33$.

\subsubsection*{Figure~\ref{fig:data}.}
Examples of experimental data and corresponding fits for DOPC liposomes.
Top Left: liposome~A; fit results are $N_c=6$, $\lambda = 0.13$ and $W_c^{fit}=0.35$.
Top Right: liposome~B; fit results are $N_c=9$, $\lambda = 0.19$ and $W_c^{fit}=0.69$.
Bottom Left: liposome~C; fit results are $N_c=1$, $\lambda = 0.15$ and $W_c^{fit}=0.51$.
Bottom Right: liposome~D; fit results are $N_c=16$, $\lambda = 0.30$ and $W_c^{fit}=0.68$.
\corrbis{Pulse magnitudes are 290, 360, 235 and 300 $V/cm$ respectively. Pulse duration is 5 $ms$. 
Arrows, if present, indicate data just before which we had to re-center the image on the liposome of interest. There is thus a time interval of $\approx$ 10 $s$ before the indicated point, instead of 2 $s$ as in all other cases.}

\subsubsection*{Figure~\ref{fig:lambda}.}
Distribution of the values of $\lambda^{fit}$ obtained after data fitting for DOPC liposomes.

\subsubsection*{Figure~\ref{fig:Wc}.}
Distribution of the values of $W_c^{fit}$ obtained after data fitting for DOPC liposomes.

\subsubsection*{Figure~\ref{fig:joli1}.}
\corr{Images} of  liposomes A, B and C, \corr{composed of DOPC and labeled with Rhodamine PE,} at times indicated by the diamonds
in Fig.~\ref{fig:data}, corresponding to 0, 12 and 24 applied pulses.
Liposomes A and C lose lipids by formation of vesicles, and liposome B
by formation of tubules.  Scalebars (20 $\mu m$ length) and positions
of the electrodes appear in the first photograph of each vesicle.
\corrbis{Pulse magnitudes are 290, 360 and 235 $V/cm$ respectively. Pulse duration is 5 $ms$. Times in upper right corners indicate when images were acquired, the time origin being the onset of the first pulse. No time indication means that the picture was taken before the first pulse.}

\begin{figure}
   \begin{center}
      \includegraphics*[width=4.8in]{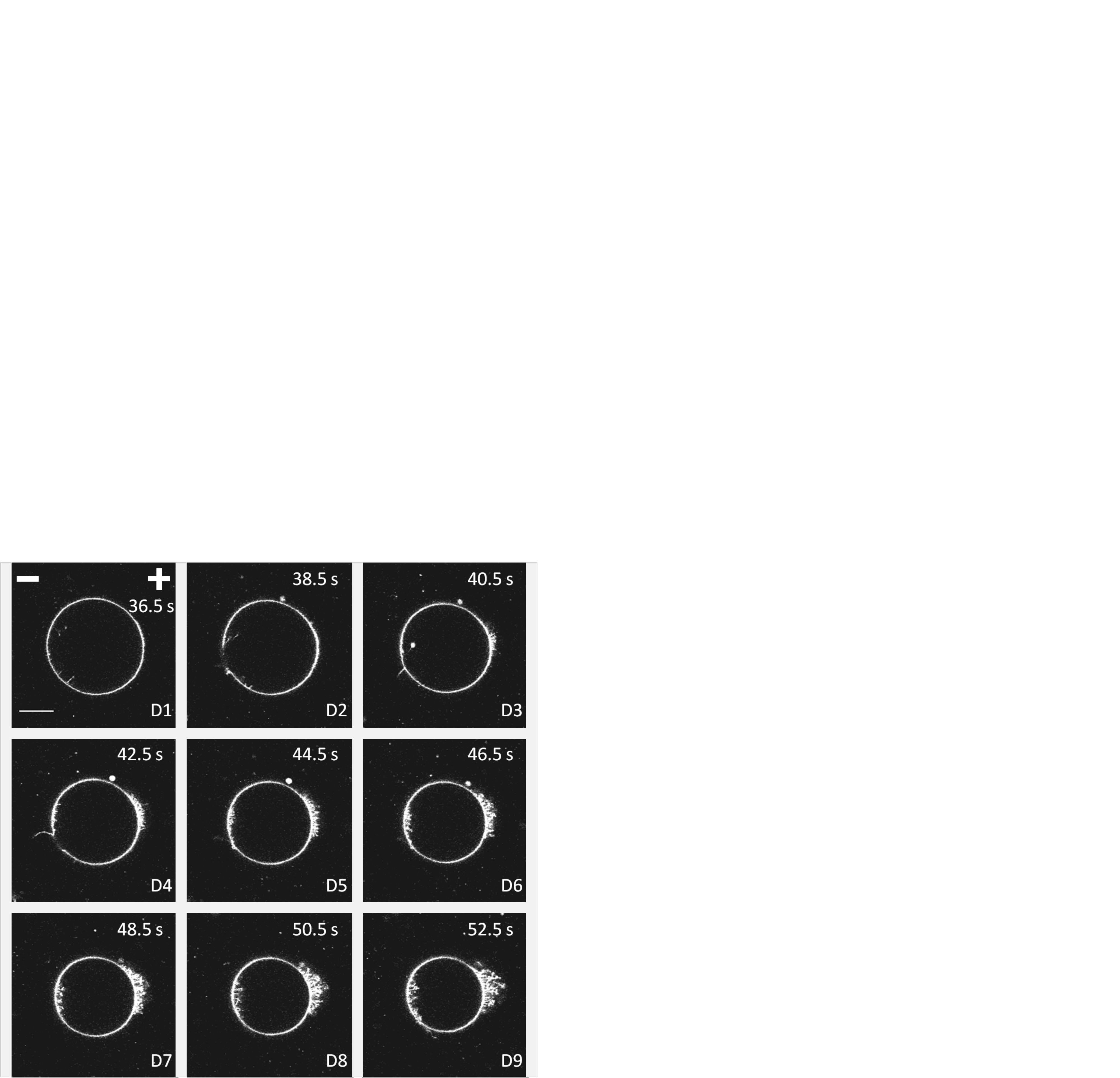}
      \caption{}
      \label{fig:joli2}
   \end{center}
\end{figure}

\subsubsection*{Figure~\ref{fig:joli2}.}
\begin{figure}
   \begin{center}
      \includegraphics*[width=4.8in]{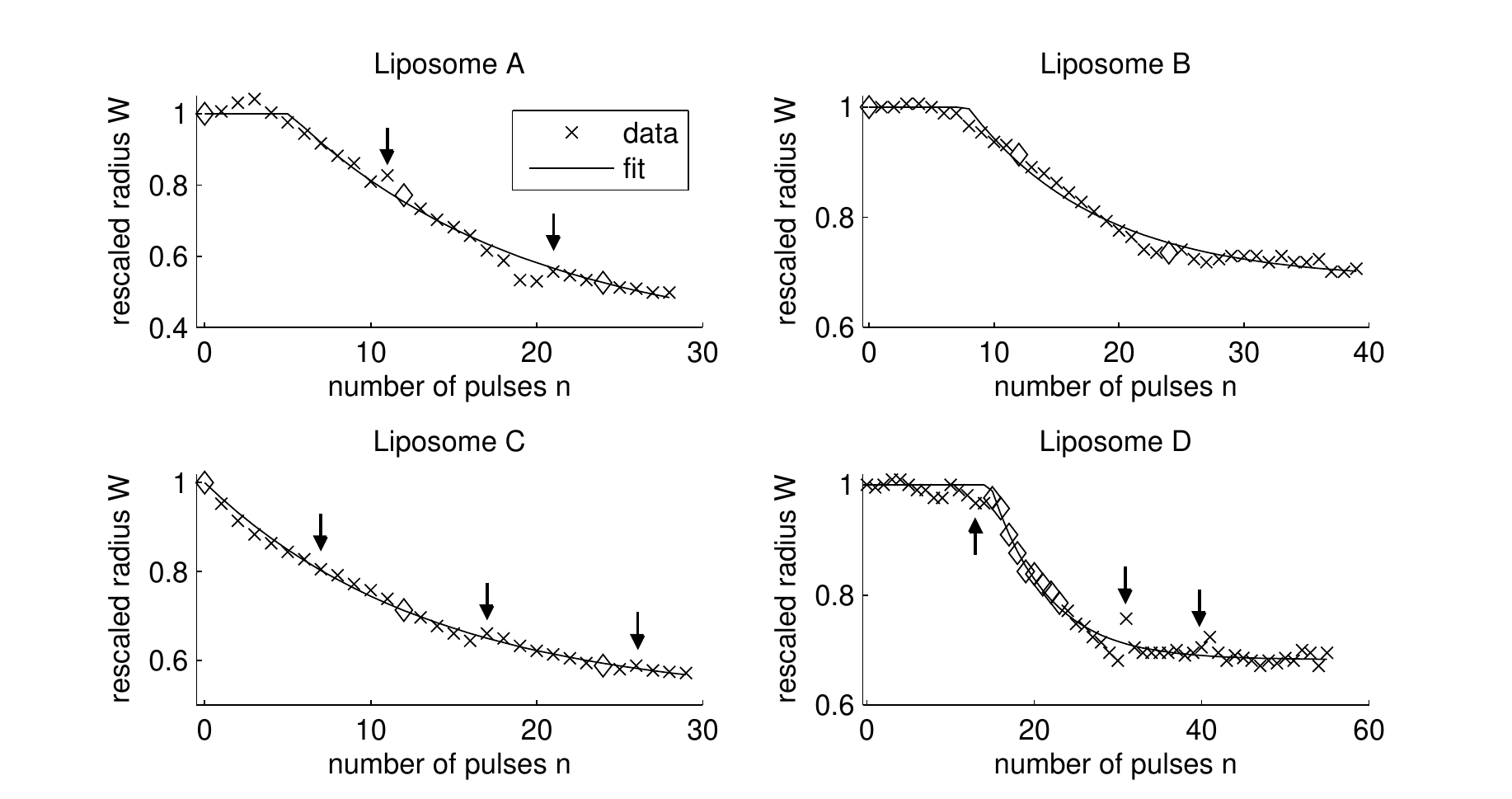}
      \caption{}
      \label{fig:data}
   \end{center}
\end{figure}

\corr{Images} of liposome D, \corr{composed of DOPC and labeled with Rhodamine PE,} at times indicated by the diamonds in
Fig.~\ref{fig:data}. Image D1 is acquired after 15 pulses, D2 after 16
pulses, D3 after 17 pulses, \latin{etc}.  We can see pores on pictures
D2 and D4 on the cathode-facing hemisphere.  Scalebar (20 $\mu m$
length) and position of the electrodes appear in the first photograph.
\corrbis{Pulse magnitude and duration are 300 $V/cm$ and 5 $ms$. Times in upper right corners indicate when images were acquired, the time origin being the onset of the first pulse.}


\clearpage

\clearpage

\clearpage

\clearpage

\clearpage

\clearpage

\end{document}